\documentclass{article}

\usepackage{color,soul}
\usepackage{graphicx}
\usepackage{hyperref}
\usepackage{subfigure}
\usepackage{amsmath}
\usepackage[numbers]{natbib}
\usepackage{tikz}
\usepackage{gensymb}
\usepackage{url}
\usepackage{hyperref}
\usepackage{booktabs}

\title{L-DIT: A dApp for Live Detectability, Identifiability and Trackability for ASOs on the Behavioral Dynamics Blockchain}
\author{\textbf{Anirban Chowdhury}$^1$, \textbf{Yasir Latif}$^{1,2}$,\\ \textbf{Moriba K. Jah}$^3$, \textbf{Samya Bagchi}$^1$\\
$^1$Space Protocol\\
$^2$Australian Institute for Machine Learning (AIML)\\
$^3$The University of Texas at Austin\\
}

\begin{document}
\maketitle

\begin{abstract}
    As the number of Anthropogenic Space Objects (ASOs) grows, there is an urgent need to ensure space safety, security, and sustainability (S3) for long-term space use. 
    Currently, no globally effective method can quantify the safety, security, and sustainability of \textit{all} ASOs in orbit. Existing methods such as the Space Sustainability Rating (SSR) rely on volunteering private information to provide sustainability ratings. However, the need for such sensitive data might prove to be a barrier to adoption for space entities.
    For effective comparison of ASOs, the rating mechanism should apply to all ASOs, even retroactively, so that the sustainability of a single ASO can be assessed holistically. Lastly, geopolitical boundaries and alignments play a crucial and limiting role in a volunteered rating system, limiting the overall safety, security, and sustainability of space.
    
    This work presents a Live Detectability, Identifiability, and Trackability (L-DIT) score through a distributed app (dApp) built on top of the Behavioral Dynamics blockchain (BDB). The BDB chain is a space situational awareness (SSA) chain that provides verified and cross-checked ASO data from multiple sources. This unique combination of consensus-based information from BDB and permissionless access to data allows the DIT scoring method presented here to be applied to all ASOs. While the underlying Behavioral Dynamics Blockchain (BDB) collects, filters, and validates SSA data from various open (and closed if available) sources, the DIT dApp presented in this work consumes the data from the chain to provide L-DIT score that can contribute towards an operator's, manufacturer's or owner's sustainability practices. Our dApp provides data for all ASOs, allowing their sustainability score to be compared against other ASOs, regardless of geopolitical alignments, providing business value to entities such as space insurance providers and enabling compliance validation and enforcement.
    
\end{abstract}

\section{Introduction}
\begin{table}[]
    \centering
    \begin{tabular}{c|c|c}   
         \textbf{Criteria} & \textbf{SSR DIT}~\cite{steindl2021developing} & \textbf{L-DIT} \\\hline
         Detectability &  Simulation &  Real Data \\\hline 
         Trackability &  Simulation & Simulation \\\hline
         Identifiability  & Angular Momentum (Sim) & Angular Momentum (Real)\\\hline
         Score gradation & Tiered & Continuous $(0,1)$ \\\hline
    \end{tabular}
    \caption{Differences between SSR DIT \cite{steindl2021developing} and the proposed L-DIT method. While trackability analysis for both methods is carried out in simulation, L-DIT can incorporate real-time ASO data for detectability and identifiability analysis and provides a scoring mechanism geared towards ranking and comparisons.}
    \label{tab:differences}
\end{table}
Rapid saturation of orbital carrying capacity (OCC) by commercial, defense, and public satellites has elevated the space community's concerns about potential collisions due to overcrowding and the resulting impact on both orbital space and our modern space-reliant economy. With 58,000 additional satellites predicted to be launched by 2030~\cite{gao_2022}, if the current orbital launch practices remain unchanged, the number of conjunctions is projected to grow exponentially.  It is therefore vital to ensure that all launches follow sustainable behavior guidelines wherein the collective operation of all ASOs leads to safer and sustainable use of space for the coming generations.

Space Safety, security, and Sustainability (S3) focuses on ensuring the long-term viability and safety of space operations and concerns itself with various aspects of space operations including space debris management, sustainability of satellite constellations, collision avoidance, and establishment of global regulatory frameworks. Currently, sustainable behavior is encouraged through guidelines such as the World Economic Forum’s (WEF) Space Industry Debris Mitigation Recommendations~\cite{WEF_space_report} or via sustainability indices such as the Space Sustainability Rating (SSR)~\cite{saada2023promoting}. While it is left to the individual actors to adopt and implement the guidelines into their practices, there is no clear path to globally enforcing sustainable practices in the current space ecosystem. However, sustainability indices can bring about change by providing economic incentives for safe and sustainable behaviors such as reducing insurance costs.

The Space Sustainability Rating (SSR) has been an initiative of the WEF’s Global Future Council on Space Technologies to encourage responsible behavior in space. The SSR aims to increase transparency of actors' debris mitigation efforts by providing a comprehensive rating for the sustainability of space missions. This rating system evaluates missions through a questionnaire, utilizing existing indices and other information to establish a rating. It serves as a voluntary mechanism where actors can demonstrate their commitment to debris mitigation without disclosing sensitive or proprietary information. 

The adoption of the SSR has been slow due to its voluntary nature. One of the limitations of the SSR is its dependency on actor-supplied sensitive information which is used as an input to provide the sustainability rating. Providing sensitive information may discourage participation from international space actors, reducing the SSR’s efficacy as a tool for effective comparison of space actors. 
In this work, we argue that a \textbf{passive metric} that relies on \textbf{publicly available data} is a better basis for forming a space safety and sustainability index. A passive metric does not require sensitive information and can be applied retroactively to existing space assets for which publicly available data can be sourced. Such a methodology has the additional benefit of allowing a broader comparison of all ASOs and quantifying their in-orbit behavior. Additionally, as the behavior of the ASO evolves, \textbf{live real-time data} can be incorporated to update the relevant rating. Such a rating mechanism is also \textbf{not bound by geopolitical boundaries}, as public and crowd-sourced information can be taken into account where the actors themselves might not be willing to volunteer sensitive data to a central rating authority. 

\section{Establishing Trust: An SSA blockchain}
How can we be sure that a given piece of information is reliable, unaltered, and has not been tampered with? Trust is a major issue in the information age not only for the space industry but for any other online interaction.  Blockchains solve the trust problem by establishing consensus among various parties, without relying on a single centralized authority. Blockchain establishes trust through its decentralized structure and the use of cryptographic techniques, including:
\begin{itemize}
    \item \textbf{Decentralization}: Unlike traditional systems that often have a central authority, blockchain is decentralized and distributed across a network of computers, known as nodes. This means that no single entity has control over the entire network, which reduces the risk of fraud and corruption.

    \item \textbf{Transparency}: Every transaction on the blockchain is recorded on a public ledger, accessible to anyone in the network. This transparency ensures that all transactions are visible and verifiable by any participant, which builds trust among users.

    \item \textbf{Immutability}: Once a transaction is confirmed and recorded in a block, it cannot be altered or deleted. This is ensured by cryptographic hash functions that secure the blocks. Any attempt to change the information in a previous block would require altering all subsequent blocks, which is computationally impractical due to the amount of work required (known as proof of work).

    \item \textbf{Consensus Mechanisms}: Blockchain uses various consensus mechanisms like Proof of Work (PoW)~\cite{nakamoto2008bitcoin}, Proof of Stake (PoS)~\cite{buterin2013ethereum}, and others to validate transactions. These mechanisms require that a majority of nodes agree on the validity of transactions before they are added to the blockchain, ensuring that no single node can make unilateral changes to the ledger.

    \item \textbf{Cryptography}: Blockchain uses cryptographic techniques such as digital signatures to ensure that transactions are securely authorized. Each participant has a pair of cryptographic keys: a public key that is openly shared and a private key that is kept secret. Transactions are signed digitally using the private key, and this signature can be verified by anyone using the corresponding public key, ensuring the authenticity of the transaction and the integrity of the sender.

\end{itemize}

\subsection{The BDB chain: Openly sourced Live-DIT rating}
In this work, we introduce an openly sourced live DIT  (L-DIT) rating system, whose primary goal is to formulate an index relying solely on live (real-time) verified publicly available data and to rank all space assets and entities. The proposed rating system is passive,  does not depend on sensitive data, takes in real-time data for analysis, and can be applied retrospectively to all ASO.  

Where does data from such an effort come from? The data is sourced from the Behavioral and Dynamics Blockchain (BDB)~\cite{BDB}. This is a Space Situational Awareness (SSA) blockchain being developed to record SSA data onto a blockchain. Different space entities including SSA providers, hobbyists, and public SSA efforts (such as SatNOGS) contribute Tracking Data Messages (TDMs) to the chain, which are cross-validated and verified on-chain, before being added to the distributed ledger as blocks. BDB therefore serves as a decentralized provider of ASOs orbital predictions (Two-Line Elements) in a distributed and decentralized manner. 

\begin{figure}
    \centering
    \includegraphics[width=\textwidth]{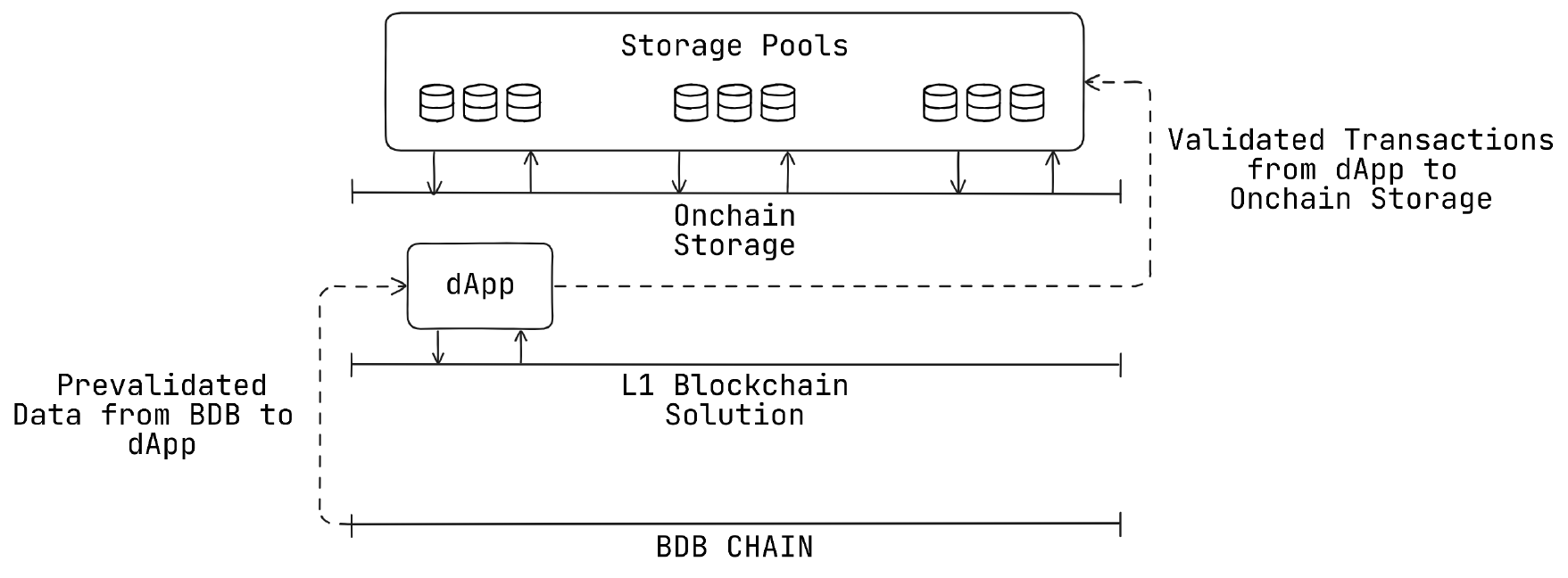}
    \caption{A distributed App (dApp) on the Behavior and Dynamics Blockchain for L-DIT.}
    \label{fig:ossr}
\end{figure}

We envision our work to reside as a distributed App (dApp) on the BDB blockchain (See Fig.\ref{fig:ossr}). The BDB chain consumes public data along with Tracking Data Messages (TDMs) from various sources including SSA providers and hobbyists, validates and cross-verifies it, and using on-chain consensus provides a single source of truth over which the analysis for O-SSR can be carried out. 

With the rise in crowd-sourced space surveillance efforts such as SatNOGs\cite{grcon} and SatObs\cite{SatObs}, more information is being cataloged by hobbyists and is made publicly available. The BDB chain consumes this information and provides a coherent view of the SSA landscape at any given moment.

While we develop BDB, the current version of the dApp presented in this work, incorporates public data from national space agencies released as a part of their public SSA efforts. 




The proposed dApp combines individual scores from the detectability (Sec. \ref{sec:detectability}), identifiability (Sec. \ref{sec:identifiability}), and trackability (Sec. \ref{sec:trackability}) modules into a single sustainability score, which ranks all ASOs in orbit around the Earth with using public information. Furthermore, individual asset scores contribute to the formulation of entity scores. Utilizing public information about ownership, operators and manufacturers for known assets, average values are computed to form the entity scores, allowing a single number to make comparisons among owners, operators, and manufacturers, expanding the horizon of what was previously not possible in terms of sustainability scores.

The current development of the rating indicates its viability as a basis of comparison for both entities and assets in space using publicly available information.  We strongly believe that such an index will guide the community in navigating the uncharted landscape of space safety and sustainability by providing a permissionless, open, and practical rating system.

\subsection{dApps: What are they?}
A decentralized app or dApp is a stack of software that typically operates automatically on a decentralized computing, blockchain network, or distributed ledger system using smart contracts. dApps can be classified into two segments: one that runs on its blockchain and one that integrates with other chains. \textbf{Smart Contracts} are used to maintain data and execute operations on the blockchain network. Complex dApps can have multiple smart contracts. Although the majority (75\%) of dApps are made of a single smart contract \cite{wu2019look}.
Consensus mechanisms are used by dApps to establish consensus on the network. There are several reasons to use a dApp including 
            decentralization,
            transparency, and security. Moreover, being distributed a dApp offers No downtime and global, uncensorable access to information.

\subsection{dApp Architecture}
The dApp is implemented of a layer-1 (L1) blockchain that supports smart contracts (See Fig.~\ref{fig:ossr}). Data from BDB flows through a bridge to the dApp when the user interacts with it to compute the rating for a particular ASO. This interaction triggers the execution of smart contracts that convert the raw ASO data into L-DIT ratings. The resulting transactions are verified by the underlying L1 chain through its consensus mechanism. At the same time, the resulting L-DIT scores are stored on long-term on-chain storage to serve as a permanent record of the ASOs rating over time. 

\section{Detectability}\label{sec:detectability}
Detectability refers to the ability of an ASO to be observed by from Earth or space-based sensing systems. The sensor is not aware of any prior information about the ASOs such as the orbit that it occupies, its mass, and other properties. Detectability solely relies on the likelihood of being observed from any of the sensors in the observational network.

Enhancing the detectability of space objects is essential for operational safety and sustainability, enabling more accurate tracking, better prediction of potential collisions, and effective avoidance maneuvers. For satellite operators and mission designers, improving detectability involves considerations around the design phase, such as incorporating materials or features that increase reflectivity or radar cross-section (RCS), and choosing orbits that optimize tracking efficacy while minimizing the risk of collision and debris generation. For the sensor end improving detectability involves better sensing capabilities and increasing the opportunities for observation by utilizing a geographically distributed network of sensors with mixed capabilities. 

Detectability of an ASO depends on:

\begin{itemize}
    \item \textbf{Size and Reflectivity}: Larger objects or those with materials that reflect radar or optical signals are easier to detect. Although the minimum size detectable depends on the sensitivity and wavelength of the tracking system, multiple observations can provide a clearer understanding of the inherent detectability of a space object.

    \item \textbf{Orbit Characteristics}: The object's orbit (e.g., altitude, inclination) influences detectability, as certain regions may be more densely monitored than others. Lower orbits may offer more frequent tracking opportunities but also present more background noise and potential for interference.
    
    \item \textbf{Observation Capabilities}: Detectability is not only a function of the space object but also depends on the network of ground- and space-based sensors that are used to make the observation. This includes the geographical distribution of sensors, their wavelength operation, and the processing capabilities to discriminate objects from background noise. As a rule of thumb, more observation using varied locations provides better detectability 

    \item  \textbf{Signal-to-Noise Ratio (SNR)}: A measure of the object's detectability against the background noise. Higher SNR values indicate a clearer detection opportunity.
\end{itemize}

Any meaningful metric of an ASO detectability should include a direct or inferred measure of the about factors. 

\subsection{Sensing for ASO detection}
ASO detectors come in various shapes and sizes and utilize a range of sensing modalities. The two most frequent types of sensors generally used are radar and optical telescopes. Both have complementary strengths. \textbf{Radar sensors} detect ASOs by bouncing radar signals off them. It's an active sensing mechanism and works as a sender-receiver pair. A transmitter illuminates the target so that a receiver can capture the energy reflected by the target. Radars can determine the position, velocity, size, and shape of an object. 
On the other hand, \textbf{optical telescopes} equipped with specialized cameras are used for detection by observing satellites passing overhead and utilizing the difference between the ASOs position against the background of stars for detection. This is a passive sensing modality and depends on the light from the sun reflected by the ASO towards the detector, limiting the time during which observations can be made. 

In this work, we use the Radar Cross Section (RCS) of an ASO as a measure of its detectability (See Sec. \ref{sec:detection_methodology} for the reasoning). 
The radar cross section (RCS) quantifies the detectability of an object by ground-based or space-based radar systems. It is defined as the hypothetical area required to intercept and re-radiate an amount of power back to the radar source that is equivalent to the power intercepted by the target. The RCS measures the target's effectiveness in reflecting radar signals to the receiver. The value of RCS is influenced by several factors including the size, shape, material composition, and the angle of incidence of the radar waves on the target. A higher RCS indicates greater detectability by radar, whereas stealth technology aims to minimize the RCS of objects. RCS is typically expressed in square meters \(\text{m}^2\) or decibels relative to a square meter (dBsm).
The RCS value in dBsm provides a standard reference point for comparing the reflectivity of different objects.

\subsection{Methodology}\label{sec:detection_methodology}

This section provides further details of the proposed detectability metric, how it is derived and the data that is used as input.

\begin{figure}[t]
  \centering
  \includegraphics[width=0.75\linewidth]{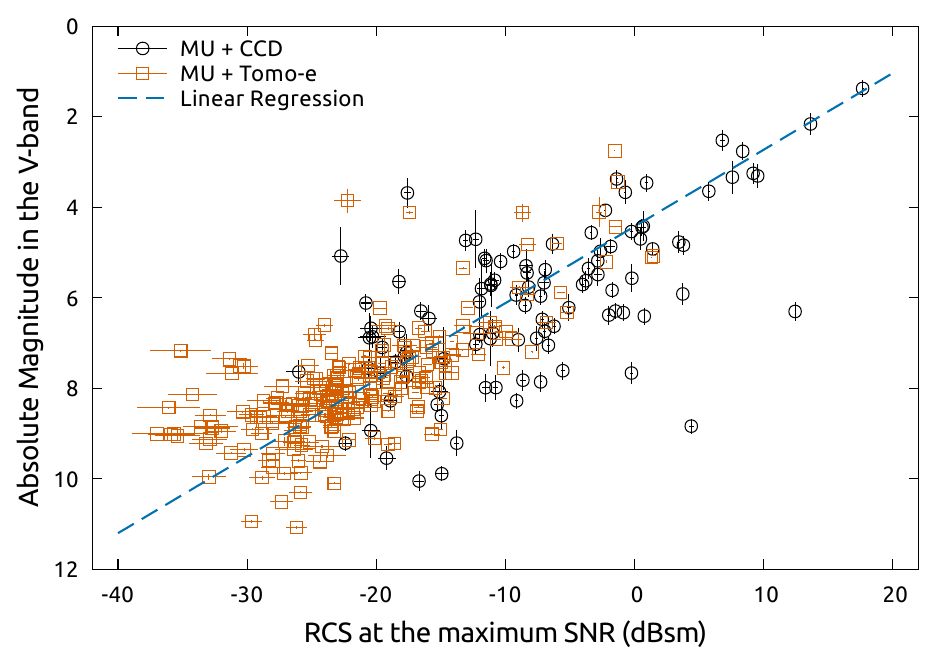}
  \caption{Relationship between RCS and Absolute Magnitude for faint meteor observations. (Figure taken from \cite{Ohsawa_2020})}
  \label{fig:rcs-magnitude}
\end{figure}

Ideally, the observations made by various sensors can been assumed to be independent of each other. However, the work in \cite{Ohsawa_2020}, studied the relationship between RCS and optical magnitude for faint meteors. They show that a linear relationship exists between the measurements made by the two sensors over a range of about 1–9 mag as shown in Fig.~\ref{fig:rcs-magnitude}.

\begin{figure}[hbt!]
      \centering
      \includegraphics[width=0.75\linewidth]{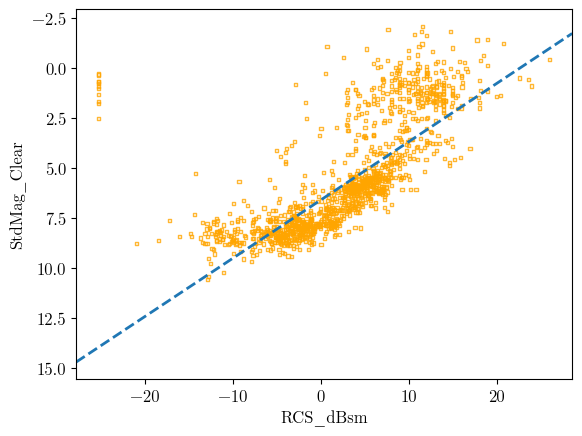}
      \caption{Relationship between RCS and standard magnitude for ASOs in orbit around the Earth.}
      \label{fig:rcsdbsm-stdmag}
\end{figure}

This provides an insight into the correlation that exists between the two sensors, higher detectability in one, is correlated to higher detectability in the other. Therefore, we consider only one sensor for detectability, the RCS, as including both can lead to bias towards more detectable ASOs. Fig. ~\ref{fig:rcsdbsm-stdmag} replicates the study in \cite{Ohsawa_2020} for ASOs closer to Earth, where a similar trend can be seen. 
For this experiment, the RCS values for ASO were sourced from Celestrack \cite{CelesTrak} and the visual magnitude data was collected from the nine-channel Mini-MegaTORTORA (MMT-9) optical wide-field monitoring system \cite{MMT9}.

\subsubsection{Data sources}
The data for the analysis will be sourced from the BDB chain, which contains verified information about ASOs, their orbits, and positions at all times. The BDB is interoperable with the two other chains in the spaceprotocol ecosystem: The Structural and Physical Properties Blockchain (SPP) and the Collaborative Operations Blockchain (COB). Together, these chains provide all the necessary information for the proposed L-DIT dApp.

For the current detectability analysis, however, we collect data from multiple sources including space-track\footnote{\url{www.space-track.org}}, CelesTrak~\cite{CelesTrak} , MMT9~\cite{MMT9} and Jonathan's Space report \cite{planet4589}. For each ASO, we collect the RCS information and where there is a conflict, we choose the most optimistic values (higher RCS). We use the normalized RCS (dBsm) as the detectability score for each ASO\footnote{The presented methodology is easy to extend to optical observations where either RCS is unavailable or unreliable. In such cases, (\ref{equ:sd}) can be extended to include visual magnitude in addition to RCS values and the maximum of the two can be considered as the final score.}. 

Given the dynamic nature of the dApp, further measurements such as radio, optical, and Satellite Laser Ranging (SLR) can be included to generate a more holistic detectability scoring mechanism.

\subsection{Results}
Given the range of RCS values for all ASOs $R = \{r_i, i = 1 \dots N\}$, we perform min-max normalization to compute the Detectability score ($S_D^i$) for the i-th ASO as 
\begin{equation}
    S_D^i = (r_i - \text{min}(R)) / (\text{max}(R)-\text{min}(R))
    \label{equ:sd}
\end{equation}
\noindent
where max$(.)$ and min$(.)$ return the maximum and minimum values. This maps all the RCS values to the range $(0,1)$, where values closer to zero signify less detectability.

\begin{figure}[t]
      \centering
      \includegraphics[width=0.75\textwidth]{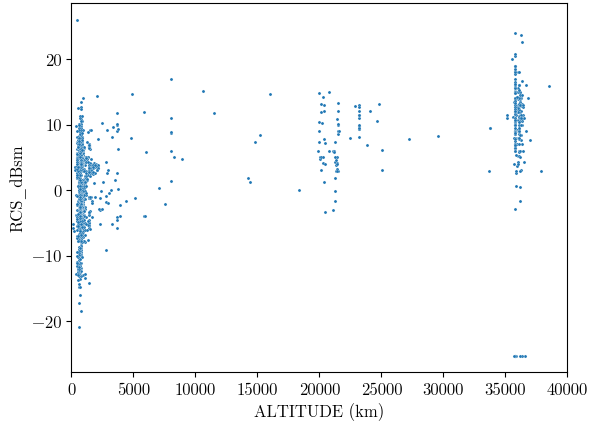}
      \caption{The RCS of ASOs at various altitudes. Higher values in GEO indicate larger ASOs present in the region.}
      \label{fig:rcs-alt}
    \end{figure}

\begin{figure}[htbp]
      \centering
      \includegraphics[width=0.75\textwidth]{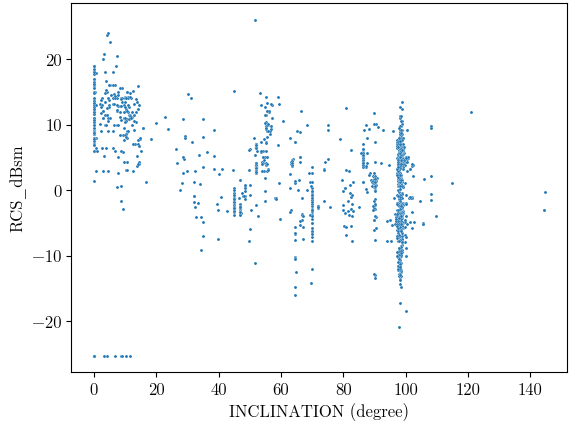}
            \caption{The RCS of ASOs at various orbit inclinations. ASOs at lower inclinations are more detectable.}
    \label{fig:rcs-incl}
    \end{figure}

\begin{figure}[htbp]
      \centering
      \includegraphics[width=0.75\textwidth]{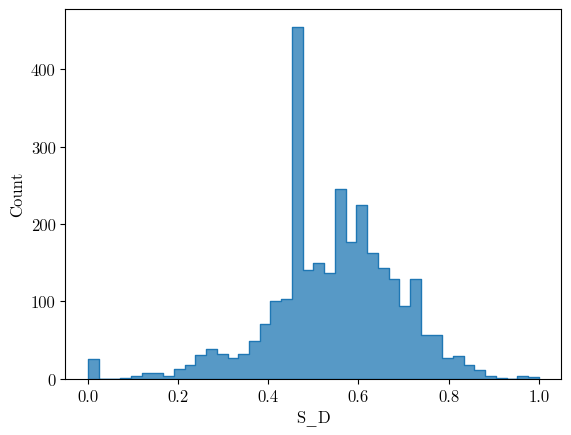}
    \caption{Distribution of detectability scores ($S_D$) for all ASOs included in the space-track with valid RCS values.}
    \label{fig:det-score}
    \end{figure}

Fig.~\ref{fig:rcs-alt} and Fig.~\ref{fig:rcs-incl} provide an overview of the RCS values at various altitudes and inclination respectively. It can be seen that larger ASOs, such as those in GEO and ASOs at lower inclination can be more easily detected. The distribution of $S_D$ for all ASOs considered is shown in Fig.~\ref{fig:det-score} and Table.~\ref{tab:det-top10} shows the ASO with the highest detectability scores.

    \begin{table}[h]
        \centering
        \begin{tabular}{@{}lc@{}}
        \toprule
        \textbf{ASO} & \textbf{$S_D$} \\ \midrule
        ECHOSTAR 3         & 1.000000           \\
        AMAZONAS 3          & 0.991693           \\
        LUCH 5B & 0.971285       \\
        NIMIQ 1            & 0.964161           \\
        ANIK G1           & 0.956798           \\
        HYLAS 1            & 0.907953           \\
        STARONE C3           & 0.903000           \\
        NIMIQ 2               & 0.901528           \\
        NIMIQ 5            & 0.887715           \\
        APSTAR 9           & 0.881620           \\ \bottomrule
        \end{tabular}
        \caption{Most detectable ASOs}
        \label{tab:det-top10}
    \end{table}
    
    \begin{table}[h]
        \centering
        \begin{tabular}{@{}lc@{}}
        \toprule
        \textbf{ASO} & \textbf{$S_D$} \\ \midrule
        GSAT 2 & 0.0 \\
        GSAT 18  0.0 \\
        STARONE D1 & 0.0 \\
        EXPRESS MD1 & 0.0 \\
        ARABSAT 4B & 0.0 \\
        APSTAR 7 & 0.0 \\
        (USA 95) & 0.0 \\
        IRNSS 1C & 0.0 \\
        ERS 29 (OV5-5) & 0.0 \\
        ERS 21 (OV5-4) & 0.0 \\ \bottomrule
        \end{tabular}
        \caption{Least detectable ASOs}
        \label{tab:det-bottom10}
    \end{table}

\section{Identifiability}\label{sec:identifiability}
Identification encompasses the process whereby an observer, lacking prior knowledge of a spacecraft's designation, ownership, orbital characteristics, and dimensions, utilizes observational data to ascertain the identity of a detected ASO. In this context of limited information, the identifiability of a spacecraft is conceptualized as the estimated challenge an observer would face in uniquely recognizing and differentiating a specific spacecraft from other orbital objects, relying solely on observational data and a general understanding of the ASO population for reference. 

Identifiability encompasses the degree to which individual space objects can be uniquely identified and distinguished from one another within the space environment. This concept is critical for effective space traffic management, collision avoidance, and the overall governance of space activities. Identifiability extends beyond the initial detection and tracking of objects to include the capability to continuously and accurately associate observed objects with specific satellites, debris, or other space assets throughout their orbital lifetime.

There are ways in which operators and manufacturers can improve the detectability of their space assets. These include

\begin{itemize}



\item \textbf{Data sharing}: Documenting and sharing data on the physical characteristics of space objects, such as size, shape, and reflectivity, aids in their identification through observational data. This can also include specific features designed to enhance detectability and trackability, like corner cube retroreflectors for optical tracking.

\item \textbf{Orbital Data}: Regular updates of an object's orbital parameters, including its current and projected trajectory, are crucial for maintaining its identifiability over time. This involves the sharing of Two-Line Element sets (TLEs) or more precise orbital data with entities such as the 18th Space Control Squadron of the United States Space Force and commercial tracking services.

\item \textbf{Active Signaling Devices}: For higher-end missions, the inclusion of active signaling devices (such as beacons that emit specific radio frequencies) can significantly enhance an object's identifiability, allowing it to be distinguished even in densely populated orbital regions.


\end{itemize}


\subsection{Methodology}
The identifiability considered in this work is passive: given a Tracking Data Message (TDM) of an observed ASO, how difficult it is to reliably associate it to a previously known ASO. This assumes that a database of known ASO is available, which is acquired via the BDB chain. 
To effectively answer whether a TDM can be uniquely assigned to an ASO, we closely follow the clustering-based techniques outlined in 
\cite{steindl2021developing}. However, the method has not been described in detail in their work. The method presented here is our vision of how clustering similar orbits should be carried out. 

Orbits come in various sizes and shapes. It is therefore important to consider a compact representation that encapsulates various aspects of the orbit of an ASO. Such a representational space can then be partitioned into clusters where each cluster contains ASOs with similar orbits.

For this, we consider the angular momentum of an ASO.
For an ASO in orbit around a central body (e.g., Earth in our case), the angular momentum $\mathbf{l}$ is given by the cross product of the ASO's position vector $\mathbf{r}$ and its linear momentum $\mathbf{p}$:
\begin{equation}
\mathbf{l} = \mathbf{r} \times \mathbf{p}    
\end{equation}

Since the linear momentum $\mathbf{p}$ is equal to the mass of the satellite $m$ times its velocity $\mathbf{v}$, we can write:
\begin{align}
    \mathbf{l}  &=  \mathbf{r} \times \mathbf{p} \\ 
     & =  \mathbf{r} \times m\mathbf{v}
\end{align}
\begin{figure}
    \centering
    \includegraphics[width=0.5\textwidth]{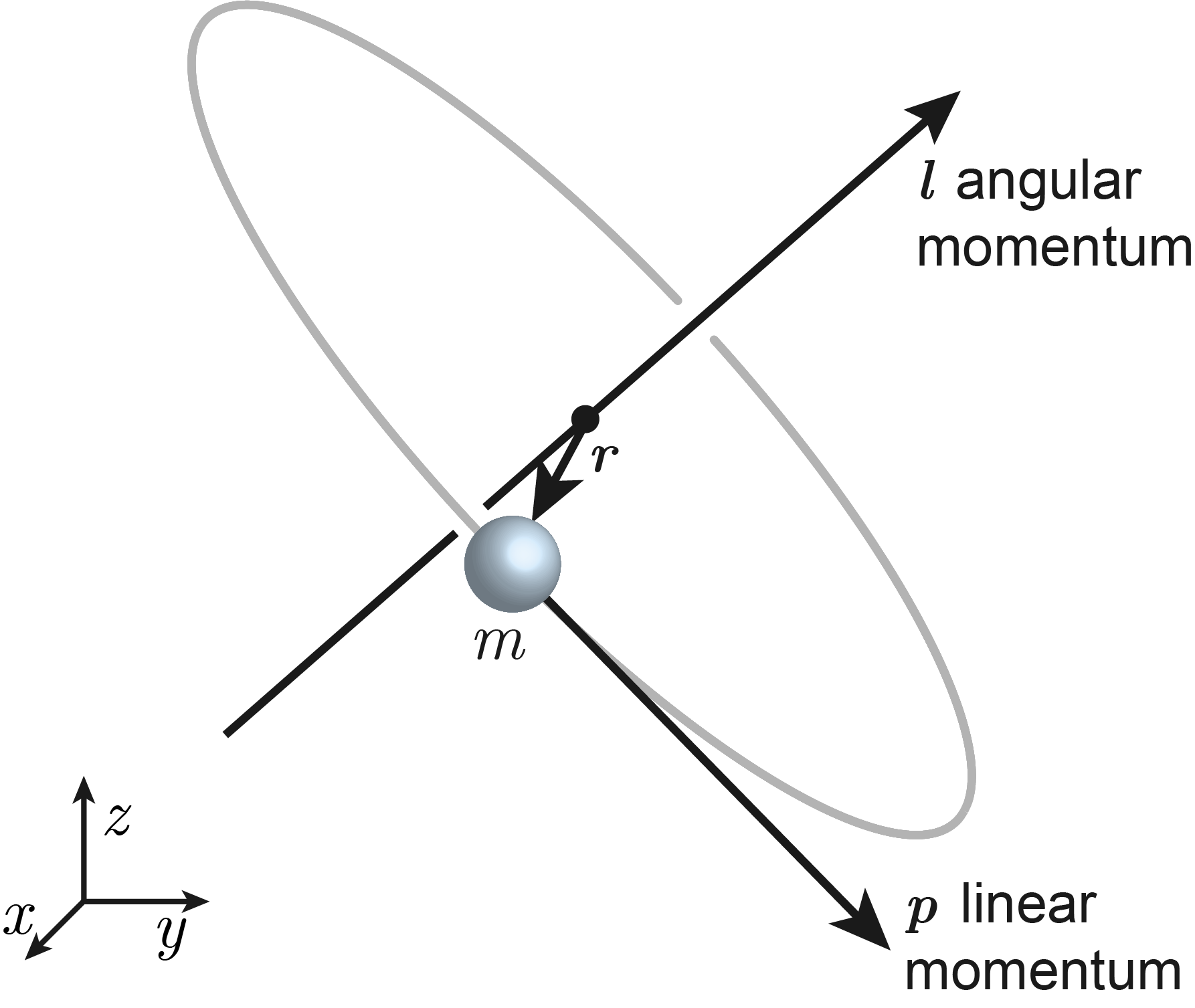}
    \caption{Angular momentum of body with mass $m$}
    \label{fig:angular-momentum}
\end{figure}
For a satellite in a circular orbit, the magnitude of the angular momentum can be simplified further. If $r$ is the radius of the orbit and $v$ is the constant orbital speed, the magnitude of the angular momentum $|l|$ is:

\begin{equation}
|l|=mrv    
\end{equation}

In the case of an elliptical orbit, the expression for angular momentum becomes more complex, as both the radius and velocity of the satellite vary along the orbit. However, the angular momentum remains constant due to the conservation of angular momentum. The direction of the angular momentum vector is perpendicular to the plane of the orbit, following the right-hand rule.

The angular momentum quantifies the various orbital parameters (velocity, mass, distance, orbital plane) conveniently into a vector $\mathbf{l} \in \mathcal{R}^3$, which can be used for clustering to find ASOs with similar orbital characteristics. To find ASOs with similar characteristics, we use bisecting k-means clustering \cite{bkm} to cluster all ASO in distinct subsets. When a new ASO is observed, its computed angular momentum is assigned to the nearest cluster center. The number of other ASOs present in the assigned cluster determines the identifiability score of the observer ASO. If the cluster size is larger, it is generally considered difficult to identify which of the ASOs in the cluster corresponds to the current observation, making it less identifiable. The smaller the size of the matched cluster, the more identifiable the ASO is considered. More specifically, the identifiability score for a cluster ($C_I$) of size $N$ is given by
\begin{equation}
    C_I = \frac{1}{\sqrt{N}+1}
\end{equation}

We compute this score for each ASO by associating it to its nearest cluster, identified by its cluster center. The final identifiability score is obtained by computing the min-max normalization over the cluster-assigned scores.
Given the set of the identifiability score $I = \{C_I^i, i = 1 \dots N\}$, for all ASO, the identifiability score for the $i$-th ASO is given by:
\begin{equation}
    S_I^i = (C_I^i - \text{min}(I)) / (\text{max}(I)-\text{min}(I))
\end{equation}
\noindent
resulting in all score to lie in the range $(0,1)$.

\subsection{Results}
\begin{figure}
    \centering
    \subfigure[]{\includegraphics[width=0.40\textwidth]{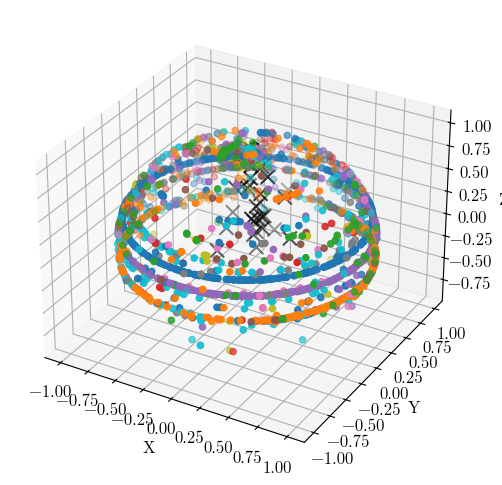}}
  \subfigure[]{\includegraphics[width=0.45\textwidth]{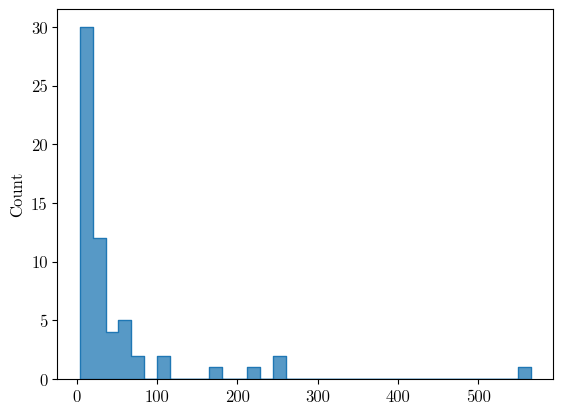}} \\
    \caption{k-means clustering in the momentum space \textbf{a:} The clusters found in the angular momentum space, shown on the hemisphere of the direction of angular moment. Each color represents a different cluster. \textbf{b:} The distribution of cluster sizes.}
    \label{fig:kmeans-clusters}
\end{figure}

\begin{figure}
    \centering
    \subfigure[Smallest cluster]{\includegraphics[width=0.45\textwidth]{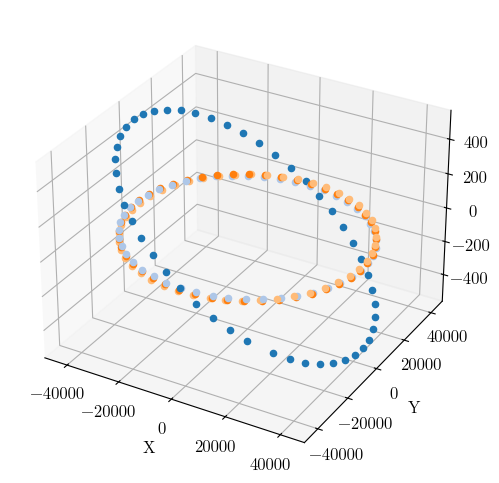}}
\subfigure[Closest clusters]{\includegraphics[width=0.45\textwidth]{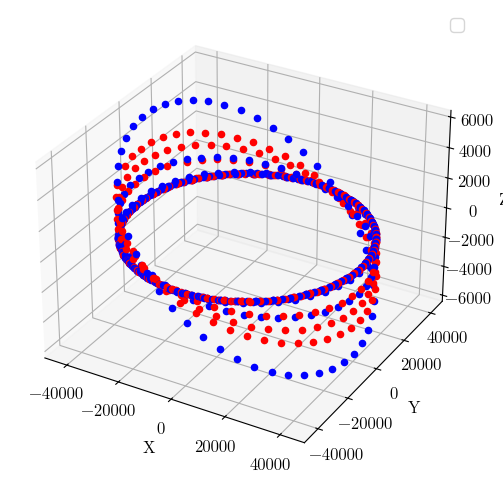}}
    \caption{A look inside the clusters: \textbf{a)} The smallest cluster discovered, which forms the most identifiable ASOs. \textbf{b)} The two closest clusters in the angular momentum space are shown in different colors. This demonstrates that clusters close to each other in the representational space have meaningful interpretations in the orbital space.} 
    \label{fig:smallest-cluster}
\end{figure}

We first present the results for the k-means clustering which shows the number of clusters (empirically found to be 60). The results are shown in Fig.~\ref{fig:kmeans-clusters} where each cluster is represented using a different color. The clustering leads to a few large clusters (low identifiability) and many smaller clusters (high identifiability). The smallest cluster contains only three orbits and are visualized in Fig.~\ref{fig:smallest-cluster}. This demonstrated the efficacy of the angular momentum space as a viable representation for clustering. 

The most and least identifiable ASO are given in Table.~\ref{tab:ident-top-10} and Table.~\ref{tab:ident-bottom-10}.
\begin{table}[htbp]
    \begin{minipage}[t]{.5\textwidth}
      \centering
      \label{tab:top10}
      \begin{tabular}{|c|c|c|}
        \hline
        \textbf{ASO} & $S_{I}$ \\
        \hline
        AMAZONAS 2 & 1.000000 \\
        CIEL-2 & 1.000000 \\
        STARONE C4 & 1.000000 \\
        ECHOSTAR 15 & 1.000000 \\
        FENGYUN 4A & 0.917103 \\
        BADR-5 (ARABSAT 5B) & 0.917103 \\
        SKYTERRA 1 & 0.917103 \\
        CHINASAT 1A & 0.917103 \\
        MORELOS 3 & 0.917103 \\
        EXPRESS AMU1 & 0.851925 \\
        \hline
      \end{tabular}
      \caption{10 most identifiable ASOs}
      \label{tab:ident-top-10}
    \end{minipage}%
    \begin{minipage}[t]{.5\textwidth}
      \centering
      \label{tab:bottom10}
      \begin{tabular}{|c|c|c|}
        \hline
        \textbf{ASO} & $S_{I}$ \\
        \hline
        A-1 (ASTERIX) & 0.0 \\
        FORMOSAT 3B & 0.0 \\
        FORMOSAT 3A & 0.0 \\
        COSMOS 1716 & 0.0 \\
        COSMOS 1717 & 0.0 \\
        COSMOS 1718 & 0.0 \\
        COSMOS 1719 & 0.0 \\
        COSMOS 1720 & 0.0 \\
        COSMOS 1721 & 0.0 \\
        FORMOSAT 3C & 0.0 \\
        \hline
      \end{tabular}
        \caption{10 least identifiable ASOs}
        \label{tab:ident-bottom-10}
    \end{minipage}
    \end{table}


\section{Trackability}\label{sec:trackability}

Trackability refers to the capability of continuously monitoring and predicting the trajectory of space objects with high precision over time. It is an essential task for space operations, contributing significantly to space traffic management and the prevention of in-orbit collisions. Trackability extends beyond mere detectability to encompass the ongoing ability to accurately follow an object's orbit and foresee its future positions.

Trackability is vital for the sustainability of space activities, enabling effective space traffic management, enhancing the safety of space operations, and reducing the risk of collisions. In the SSR framework, missions that demonstrate robust trackability through their design, operational strategies, and adherence to best practices in data sharing and cooperation are likely to achieve a higher rating, signifying a greater contribution to the long-term sustainability of the space environment. In contrast, the current version of the dApp implements trackability in terms of the observation of ASOs from a set of ground-based sensors.

    
\subsection{Methodology}
\begin{figure}[t]
  \centering
  \includegraphics[width=\linewidth]{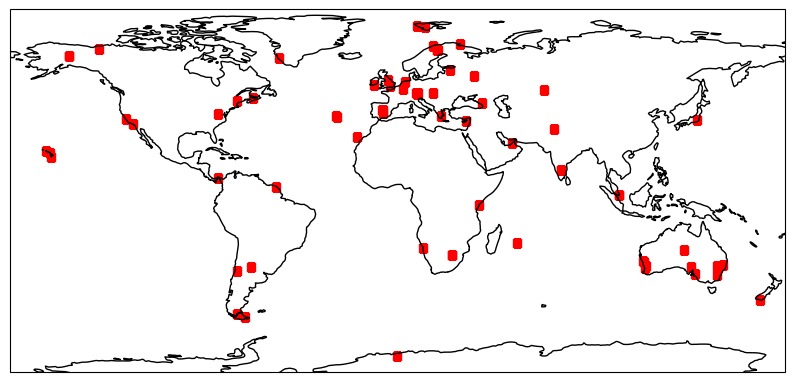}
  \caption{Locations of Ground Stations used in the Monte Carlo Simulation for trackability analysis}
  \label{fig:ground_stations}
\end{figure}

Trackability is a function of the observing sensor network as it enables the opportunities for the detection of ASOs. We therefore formulate the trackability of an ASO based on how frequently and for how long it is observed by a simulated sensor network (See Fig.~\ref{fig:ground_stations} for locations). To simulate noise and intermittent failures of sensors, we compute the detectability score using a Monte Carlo simulation. Monte Carlo simulation is a computational technique that uses random sampling and statistical modeling to estimate the possible outcomes of a complex system.  The choice of distribution can impact the simulation's accuracy, bias, variance, and the level of detail captured.
        \begin{itemize}
            \item The dependence on sensor distribution refers to how the placement or arrangement of these sensors affects the behavior and characteristics of the simulated system.
            \item The sensors or the array of sensors distributed globally aren't always available for tracking an object in space, various objects are being tracked across various locations at the same/ different times of the day.
            \item Taking a subset of locations for simulating the passes to calculate various parameters and then taking their average across several runs gives us a better representation of the real-time system, tracking objects across multiple locations.
        \end{itemize}
The trackability of a satellite is determined using three parameters - 
                \begin{itemize}
                    \item \textbf{The average pass duration} of a satellite is the average amount of time a satellite is visible during each orbit as it passes over a particular location on the Earth's surface. This duration is the difference between the rise and the set time. In this case, we used this as a criterion to determine the average amount of time it is visible during each pass across several different locations on Earth chosen randomly from a list of actual positions of ground stations. 
                    \item \textbf{The average interval duration} for a satellite refers to the average time between consecutive passes over a location or how frequently a satellite revisits that particular location on Earth. A shorter interval means frequent revisits. To keep things consistent, we have inverted the values after normalizing them to make the values close to one represent more frequent passes over a particular location.
                    \item \textbf{The average ground coverage} during a satellite pass is the percentage of ground stations from where it can be detected or tracked. To simulate a dynamic ground station network, a subset of locations was used within a Monte Carlo simulation, during each trial. \end{itemize}    
                    
\par The scores obtained are min-max normalized to get a value within the range of 0 to 1, with 0 representing the lowest and 1 representing the highest.
                    \[ D_T =  \frac{1}{3} \left( \text{AvgPassDur} + \text{AvgIntDur} + \text{AvgCov} \right) \]

When discussing satellites and their pass times in terms of ``rise'' and ``set'', we are typically referring to how they appear from the perspective of an observer on the Earth's surface.
\begin{itemize}
    \item Satellite Rise and Set:
    \begin{itemize}
        \item Rise: The moment when a satellite becomes visible above the horizon as it moves along its orbital path. This is when it transitions from being below the observer's horizon to being visible in the sky.
        \item Set: The moment when a satellite disappears below the horizon as it continues along its orbit. This is when it transitions from being visible in the sky to being obstructed by the Earth's curvature.
    \end{itemize}
    \item Factors Affecting Rise/Set Time:
    \begin{itemize}
        \item Orbital Inclination: Satellites with higher inclinations will appear to rise and set more towards the poles for observers at mid-latitudes.
        \item Observer's Latitude: Observers closer to the poles may experience longer periods of visibility (continuous pass) for polar orbiting satellites.
        \item Satellite Altitude: Higher altitude satellites will appear to move more slowly across the sky, affecting the duration between rise and set. \par
    \end{itemize}
\end{itemize} 

Satellites in GEO do not have a 'rise' and 'set' time like LEO/MEO satellites. They are positioned to appear stationary in the sky relative to an observer's location on the ground. This steady position allows for continuous visibility and is advantageous for applications requiring constant coverage over a specific area. Unlike geostationary satellites, geosynchronous satellites may have slight north-south motion as they orbit the Earth.

\subsection{Results}
\begin{figure}[htbp]
  \centering
  \includegraphics[width=0.75\textwidth]{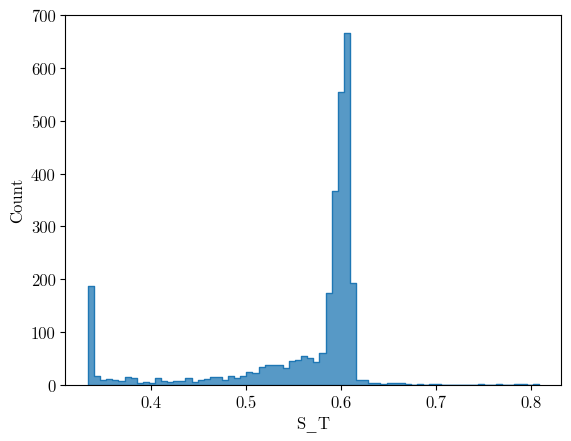}
  \caption{Distribution of Trackability score ($S_T$) for ASO included in the study}
  \label{fig:track-dist}
\end{figure}

 The 10 most trackable ASO are given in Table.~\ref{tab:track-top10} and the distribution for all ASOs is shown in Fig.~\ref{fig:track-dist}

\begin{table}[htbp]
\centering
\label{tab:object_scores}
\begin{tabular}{@{}lc@{}}
\toprule
\textbf{ASO} & \textbf{$S_T$} \\ \midrule
THEMIS D              & 0.808366            \\
CLUSTER II-FM8 & 0.795446            \\
CLUSTER II-FM7                & 0.795337            \\
CLUSTER II-FM5              & 0.784676           \\
CLUSTER II-FM6     & 0.769143            \\
DELTA 4 DEMO SPACECRAFT     & 0.748091            \\
CCE 1 (AMPTE)             & 0.745659            \\
ISO              & 0.704864            \\
THEMIS E              & 0.702617            \\
AMSAT OSCAR 40              & 0.695677            \\ \bottomrule
\end{tabular}
\caption{10 Most trackable ASOs}
\label{tab:track-top10}
\end{table}

\begin{table}[htbp]
\centering
\label{tab:object_scores}
\begin{tabular}{@{}lc@{}}
\toprule
\textbf{ASO} & \textbf{$S_T$} \\ \midrule
TURKMENALEM 52E/MONACOSA & 0.333333 \\
TJS-1 & 0.333333 \\
TIANLIAN 1-04 & 0.333333 \\
TIANLIAN 1-03 & 0.333333 \\
IPSTAR 1 & 0.333333 \\
DODGE 1 & 0.333333 \\
INSAT 4A & 0.333333 \\
ERS 26 (OV5-6) & 0.333333 \\
THOR 7 & 0.333333 \\
THOR 6 & 0.333333 \\ \bottomrule
\end{tabular}
\caption{10 Least trackable ASOs}
\label{tab:track-bottom10}
\end{table}


\section{Results for the L-DIT metric}

\subsection{L-DIT ranking}
    \begin{table}[t]
    \begin{minipage}[t]{.5\textwidth}
      \centering
      
      \begin{tabular}{|c|c|c|}
        \hline
        \textbf{ASO} & $S_{DIT}$ \\
        \hline
        SUPERBIRD 4 R & 0.743309 \\
        NAHUEL 1A & 0.708885 \\
        AMAZONAS 3 8 & 0.707831 \\
        THOR 7 9 & 0.704074 \\
        STARONE C4 & 0.700305 \\
        ECHOSTAR 3 & 0.696665 \\
        APSTAR 6 5 & 0.692965 \\
        METOP-B & 0.691118 \\
        ALOS & 0.689107 \\
        TURKSAT 1B & 0.688892 \\
        \hline
      \end{tabular}
      \caption{The highest rated ASO according to $S_{DIT}$}
      \label{tab:dit-top10}
    \end{minipage}%
    \begin{minipage}[t]{.5\textwidth}
      \centering
      
      \begin{tabular}{|c|c|c|}
        \hline
        \textbf{ASO} & $S_{DIT}$ \\
        \hline
        ERS 20 (OV5-3) & 0.222399 \\
        XW-2F & 0.208445 \\
        XW-2E 8 & 0.205723 \\
        ERS 21 (OV5-4) & 0.203051 \\
        OPS 9325 (IDSCS 12) & 0.198872 \\
        OPS 9342 (IDSCS 20) & 0.164529 \\
        OPS 9344 (IDSCS 22) & 0.159569 \\
        OV5-9 & 0.131740 \\
        ERS 26 (OV5-6) & 0.131740 \\
        ERS 29 (OV5-5) & 0.131740 \\
        \hline
      \end{tabular}
        \caption{The least rated ASO according to $S_{DIT}$}
        \label{tab:dit-bottom10}
    \end{minipage}
    \end{table}

\begin{figure}
    \centering
    \subfigure[LEO: Two satellites with similar L-DIT scores but with different coverage, due to inclination.]{\includegraphics[width=0.5\textwidth]{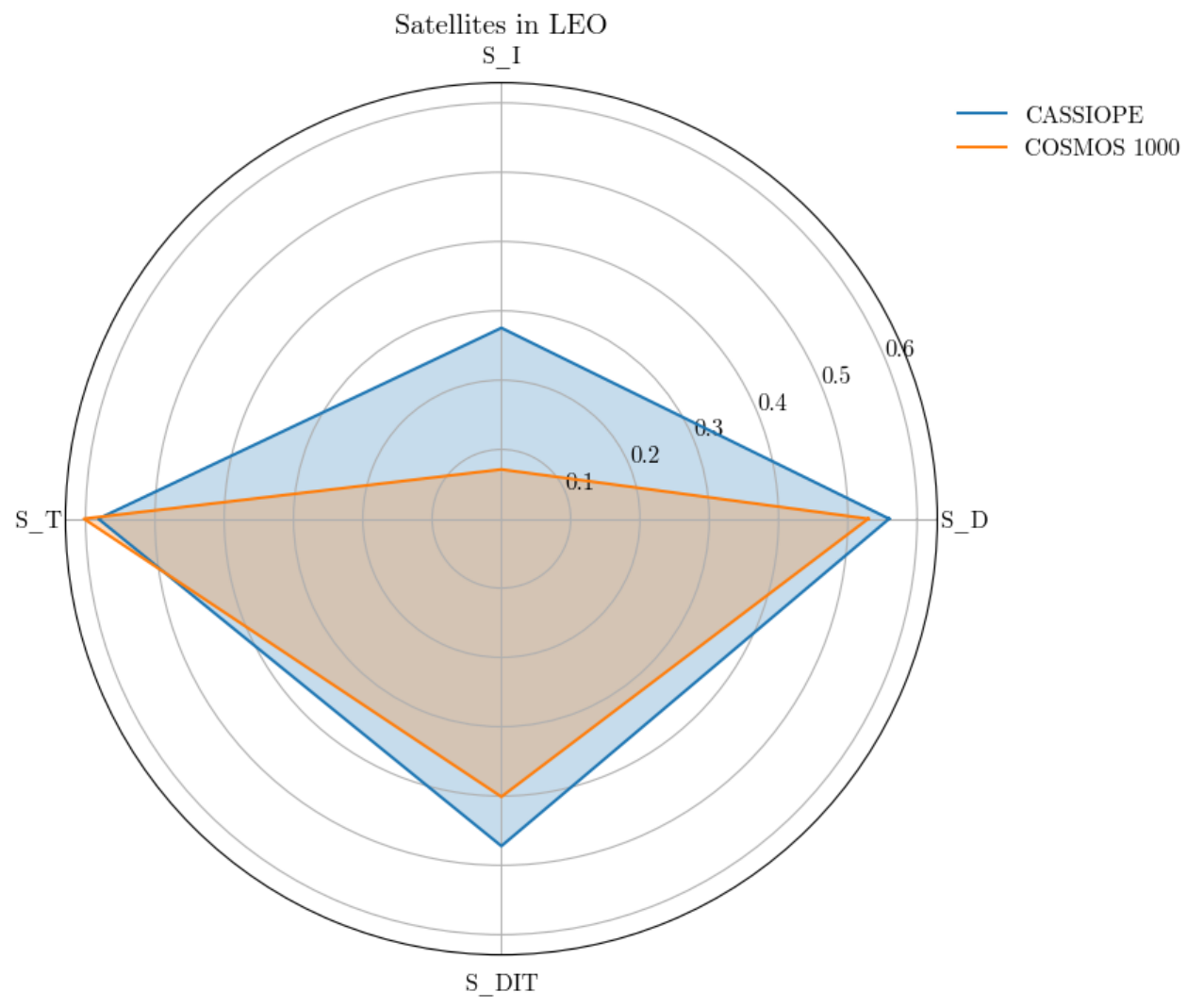}}
    \\
    \subfigure[MEO: More variation in L-DIT scores for MEO]{\includegraphics[width=0.5\textwidth]{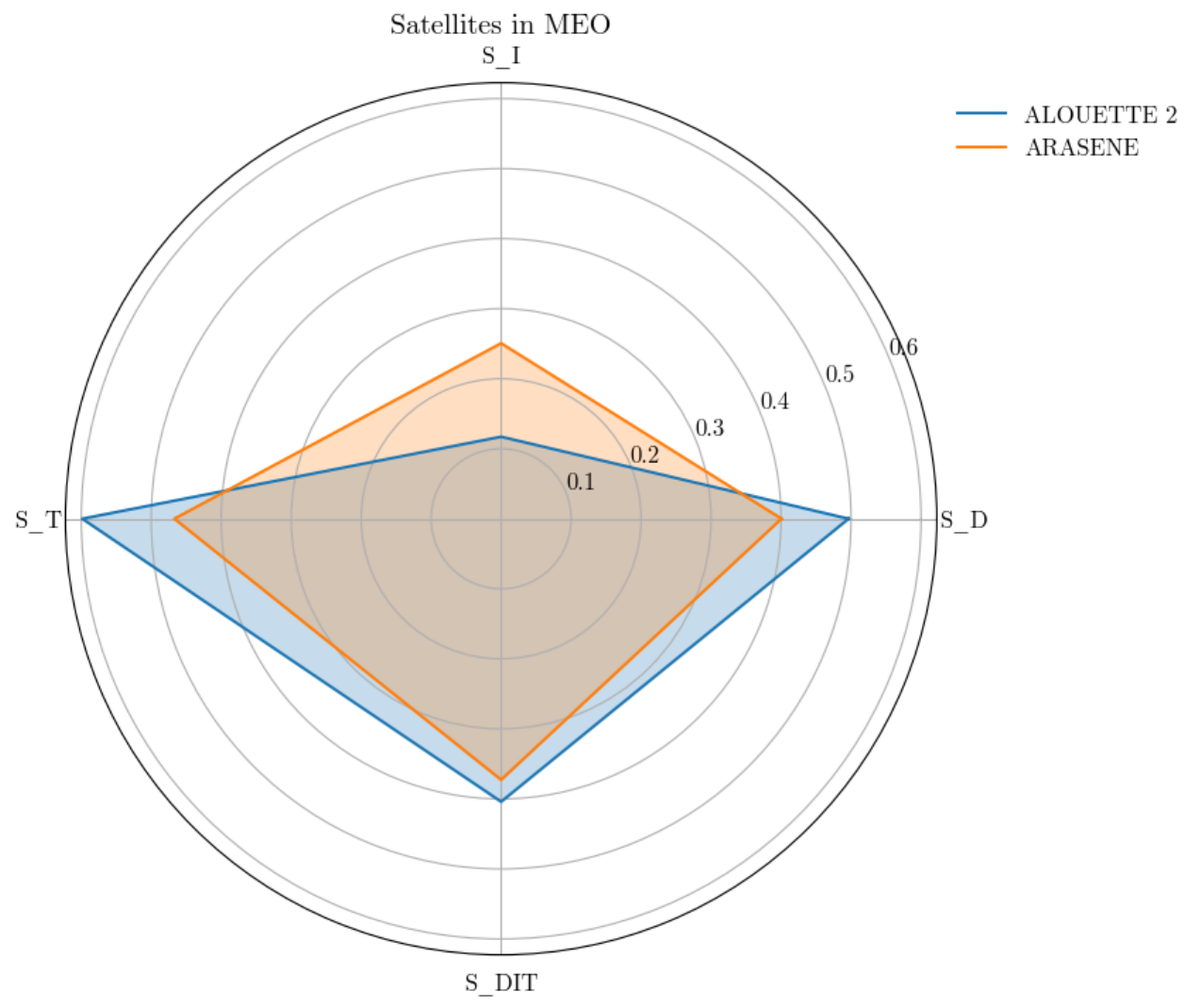}}
    \\
    \subfigure[GEO: The coverage for GEO on average is small due to being observable through only a single observer in the network.]{\includegraphics[width=0.5\textwidth]{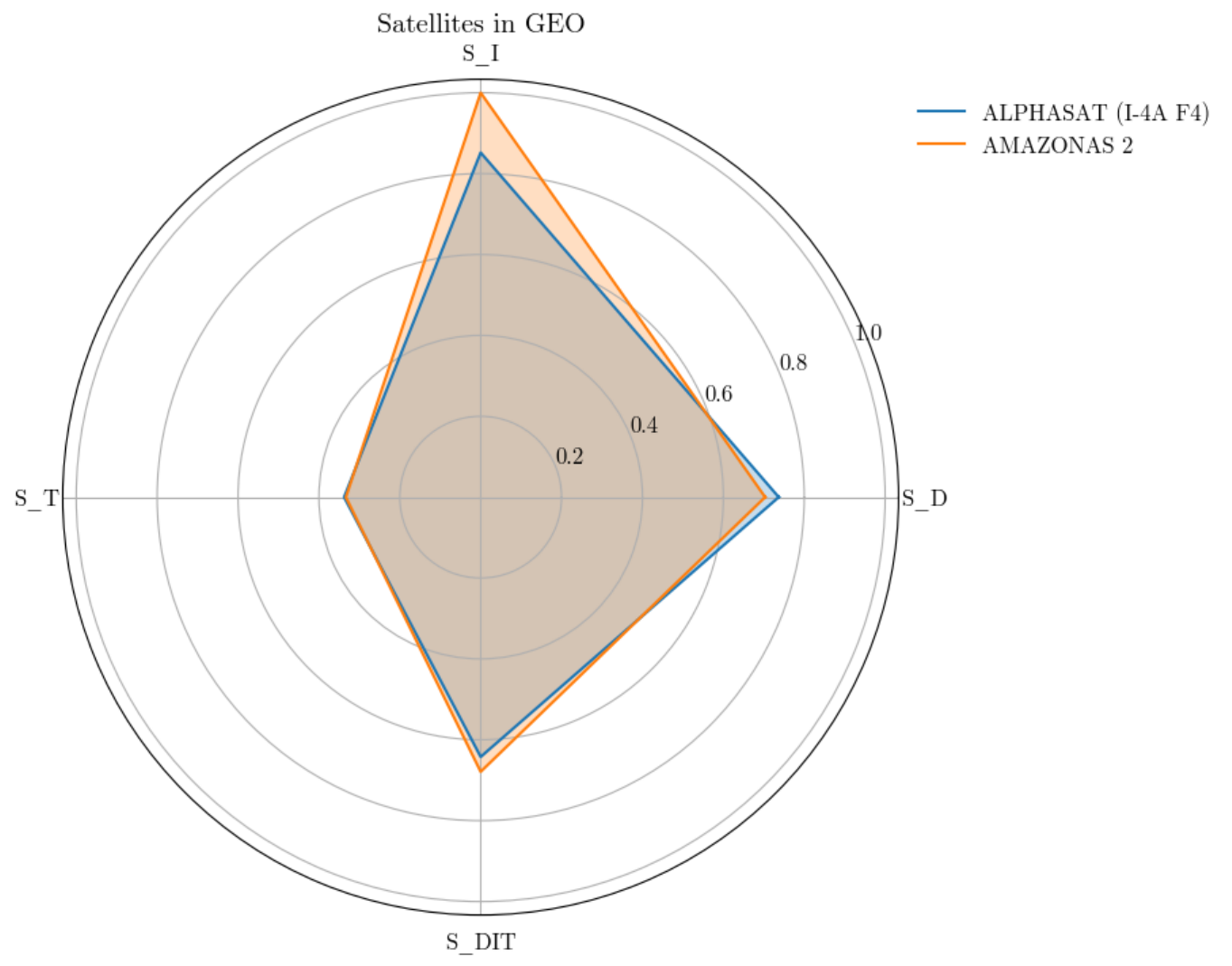}}
    \caption{Example results for the L-DIT scoring mechanism}
    \label{fig:grid}
\end{figure}

To establish the L-DIT score $S_{DIT}$ we take a mean over the detectability $S_D$, identifiability $S_I$ and trackability $S_T$ score of a satellite.
\begin{equation}
    S_{DIT} = (S_D + S_I + S_T)/3
\end{equation}
\noindent
which equally weight the three computed metrics and provide the final score for each ASO. We present representative results for combined detectability, identifiability, and trackability score in Fig.~\ref{fig:grid} where ASOs in various orbital bands are compared against each other as spider plots. An ideal satellite would appear as a diamond shape in these plots, touching the maximum value of 1 along each axis. Additionally, Table.~\ref{tab:dit-top10} and~\ref{tab:dit-bottom10} show the ASOs that score the highest and the lowest on the proposed $S_{DIT}$ score.

\section{Conclusion and Future Work}
In this work, we have proposed a method of constructing a Detectability, Identifiability, and Trackability score based on analyzing openly available data. This processing happens via a dApp on the BDB chain. The method for computation of each score is presented along with the corresponding results. The individual metrics are combined into a unified score, allowing comparisons of all ASOs currently in orbit around the Earth.

The current metric is constructed using public data only. However, actors might be interested in supplying closed-source sensitive data to further improve their ratings and provide a more comprehensive score over the baseline presented in this work. In such cases, encryption methods will be explored that can provide computations over encrypted data to securely incorporate sensitive data in the analysis. Moreover, incorporating further data sources such as SatNOGS and SeeSat-L is a possible direction for future work.

\bibliographystyle{plainnat}
\bibliography{references}
\end{document}